\documentclass[a4paper]{jpconf}
\usepackage{graphicx}
\usepackage{amsmath}
\usepackage{lineno}

\begin{document}


\title{Elliptic ($v_2$) and triangular ($v_3$) anisotropic flow of identified hadrons from the STAR Beam Energy Scan program}
\author{P.~Parfenov (for the STAR Collaboration)}

\address{National Research Nuclear University MEPhI (Moscow Engineering Physics Institute),
  Kashirskoe highway 31, Moscow, 115409, Russia}
\ead{PEParfenov@mephi.ru }

\begin{abstract}

  Elliptic ($v_2$) and triangular ($v_3$) anisotropic flow coefficients for inclusive and identified
  charged hadrons (~$\pi^\pm$, $K^\pm$, $p$, $\bar{p}$~) at midrapidity in Au+Au collisions,
  measured by the STAR experiment in the Beam Energy Scan (BES) at the Relativistic Heavy Ion Collider
  at $\sqrt{s_{NN}}$ = $11.5$ - $62.4$~GeV, are presented. We observe that the triangular flow
  signal ($v_3$) of identified hadrons exhibits similar trends as first observed for $v_2$ in
  Au+Au collisions, i.e. (i) mass  ordering at low transverse momenta, $p_T < 2$ GeV/c, (ii)
   meson/baryon splitting at intermediate $p_T$, $2< p_T < 4$ GeV/c,  and  (iii) difference in flow
   signal of protons and antiprotons.  New measurements of $v_3$ excitation function  
   could serve as constraints to test  different models and to aid new information about the
   temperature dependence of the transport properties of the strongly interacting matter.
\end{abstract}

\section{Introduction}

The heavy-ion experiments at the Relativistic Heavy Ion Collider (RHIC) and the 
Large Hadron Collider (LHC) have established the existence of a strongly coupled
Quark Gluon Plasma \cite{starQM,phenixQM}, a new state of QCD matter with partonic degrees of
freedom  and with low specific shear viscosity $\eta/s$ \cite{bass}.
Lattice QCD calculations \cite{lqcd} indicate
that the quark-hadron transition is a smooth crossover at top RHIC energy and above (
at high temperatures T and small values of baryonic chemical potential $\mu_{B}$).
A Beam Energy Scan (BES) program at RHIC plays a central role in the experimental study
of the QCD phase diagram over a wide range in T and $\mu_{B}$ \cite{starbes1,starbes2}.\\
The anisotropic flow is one of the important observables sensitive to
the equation of state (EOS) and transport properties of the strongly
interacting matter such as the shear viscosity over entropy ratio $\eta/s$ \cite{bass,vol3,lhc2}.
 The azimuthal anisotropy of produced particles can be quantified by the 
Fourier coefficients $v_n$ in the expansion of the particles azimuthal distribution 
as: $dN/d\phi \propto 1 + \sum_{n=1} 2 v_{n} \cos 
(n(\phi-\Psi_{n}))$~\cite{vol3,vol1}, where $n$ is the order of 
the harmonic, $\phi$ is the azimuthal angle of particles for a given type, 
and $\Psi_n$ is the azimuthal angle of the $n$th-order event plane. The $n^{\mathrm{th}}$-order flow
coefficients $v_n$ can be calculated as $v_{n} = \langle{\cos[n(\varphi - \Psi_n)]}\rangle$,
where the brackets denote an average over  particles and events. Elliptic  ($v_2$) and
triangular ($v_3$) flows are  the dominant flow signals and have been studied very extensively both
at top  RHIC  and LHC energies \cite{v2v3phenix,v2v3star,v2v3alice}.
For low transverse momentum ($p_T <~ $2-3 GeV/c), the $p_T$ dependence of $v_2$ and $v_3$ for
produced particles is well described by
viscous hydrodynamic models and a good agreement between data and model
calculations can be reached for 
the small values of $\eta/s$ close to the lower conjectured bound of ${1}/{4\pi}$ \cite{bass}.
The shear viscosity suppresses triangular flow signal  $v_{3}$ 
more strongly than elliptic flow signal $v_2$ \cite{v3hydro,v3hybrid}. The data for  top RHIC
energy show that, for a given collision centrality, the  measured values of $v_n$ ($n=2,3$) for  all hadrons scale 
to a single curve when plotted as $v_n/n_q^{n/2}$ versus scaled transverse kinetic energy, ($m_T-m_0$)/$n_q$, where $n_q$
is the number of constituent quarks in the hadron and $m_0$ is mass \cite{v2v3phenix,v2v3star}. 
The observed empirical Number-of-Constituent Quark (NCQ) scaling with transverse kinetic energy
may indicate that the bulk of the anisotropic flow at top RHIC energies
is partonic, rather than hadronic \cite{v2D}. The collision energy dependence of
elliptic flow ($v_2$)  for inclusive and identified hadrons at mid-rapidity
in Au+Au collisions, has been studied very extensively by STAR experiment
at $\sqrt{s_{NN}}$ = 7.7 - 62.4 GeV \cite{v2bes1,v2bes2,v2bes3,v2bes4}.
The elliptic flow signal $v_2(p_T)$  for inclusive charged hadrons
shows a very small change over such a wide range of collision energies \cite{v2bes1}. 
Hybrid model calculations show that the weak dependence of $v_2(p_T)$ on the beam energy may result from the interplay
of the hydrodynamic and hadronic transport phase \cite{v3hybrid}. The triangular flow $v_3$ is expected to be
more sensitive to the viscous damping and might be an ideal observable to probe
the formation of a QGP at different collision energies \cite{v3bes}. However, 
a significant difference in the $v_2$ values between particles and the corresponding anti-particles was
observed \cite{v2bes3,v2bes4}. This difference increases with decreasing
collision energy and is larger for baryons than mesons. Several different
theoretical models have been proposed for the possible physical reason for this effect
and new measurements of $v_3$ for particles and anti-particles
might be important for distinguishing between them \cite{v2bes3,v2bes4}. 
In this work, we report new measurements of
triangular ($v_3$) anisotropic flow coefficients for inclusive and identified
charged hadrons (~$\pi^\pm$, $K^\pm$, $p$, $\bar{p}$~) at midrapidity in Au+Au collisions at
$\sqrt{s_{NN}}$ = 11.5 - 62.4~GeV and compare them to $v_2$ results.

\section{Data Analysis}

The data reported in this analysis are from Au+Au collisions at $\sqrt{s_{NN}}$ = 11.5, 14.5, 19.6, 27, 39 and 62.4~GeV,
collected during the beam energy scan phase-I and II (BES-I \& BES-II)
programs by the STAR detector using a minimum bias trigger.  The collision vertices were reconstructed using
charged-particle tracks measured in the Time Projection Chamber (TPC). 
The TPC covers the full azimuth and has a pseudorapidity range of $\mathrm{|\eta|}<1.0$.  
\begin{figure}[h]
  \vspace{-1.5pt}
	\centering
	\includegraphics[width=0.4\textwidth]{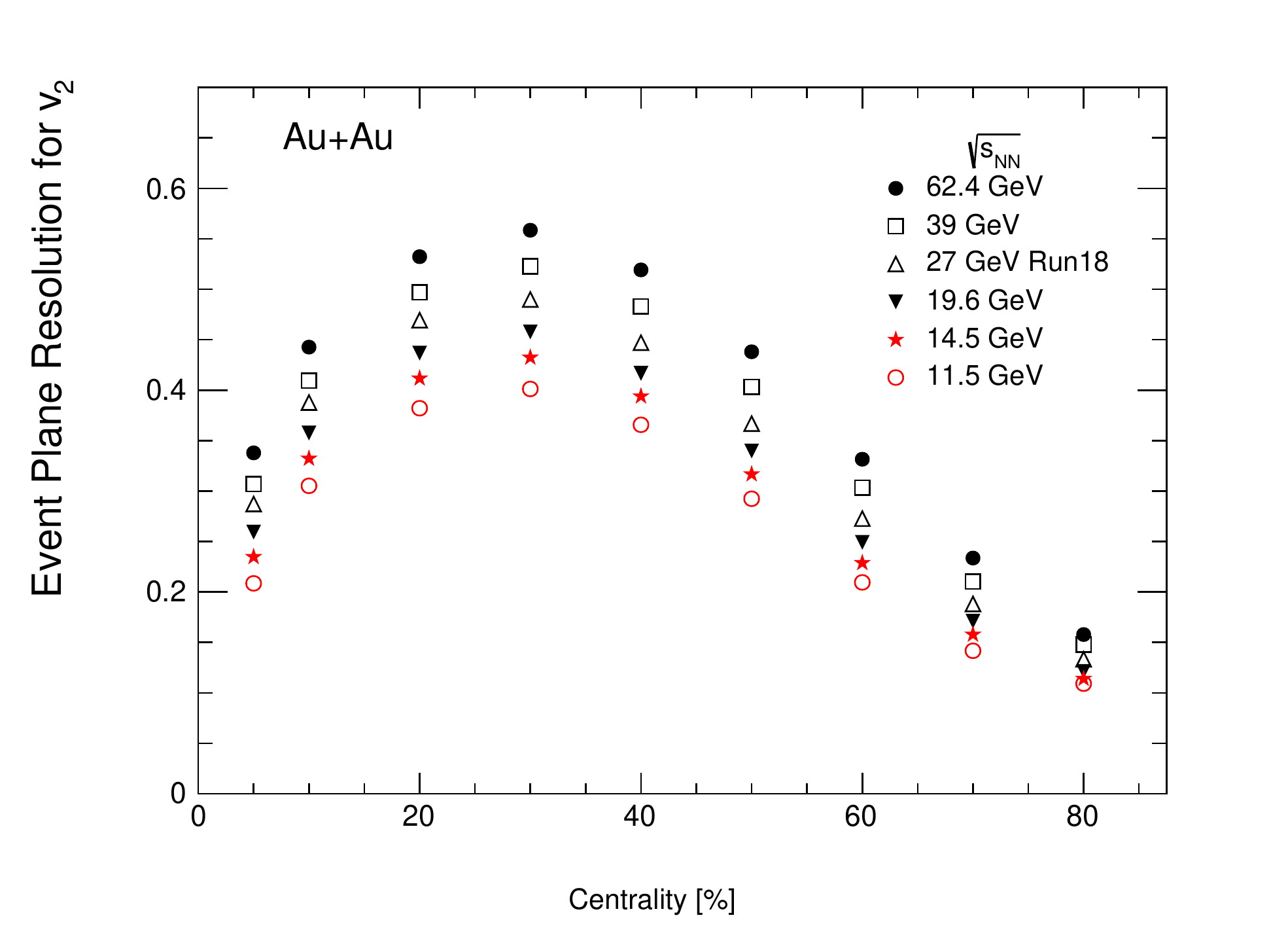}
	\includegraphics[width=0.4\textwidth]{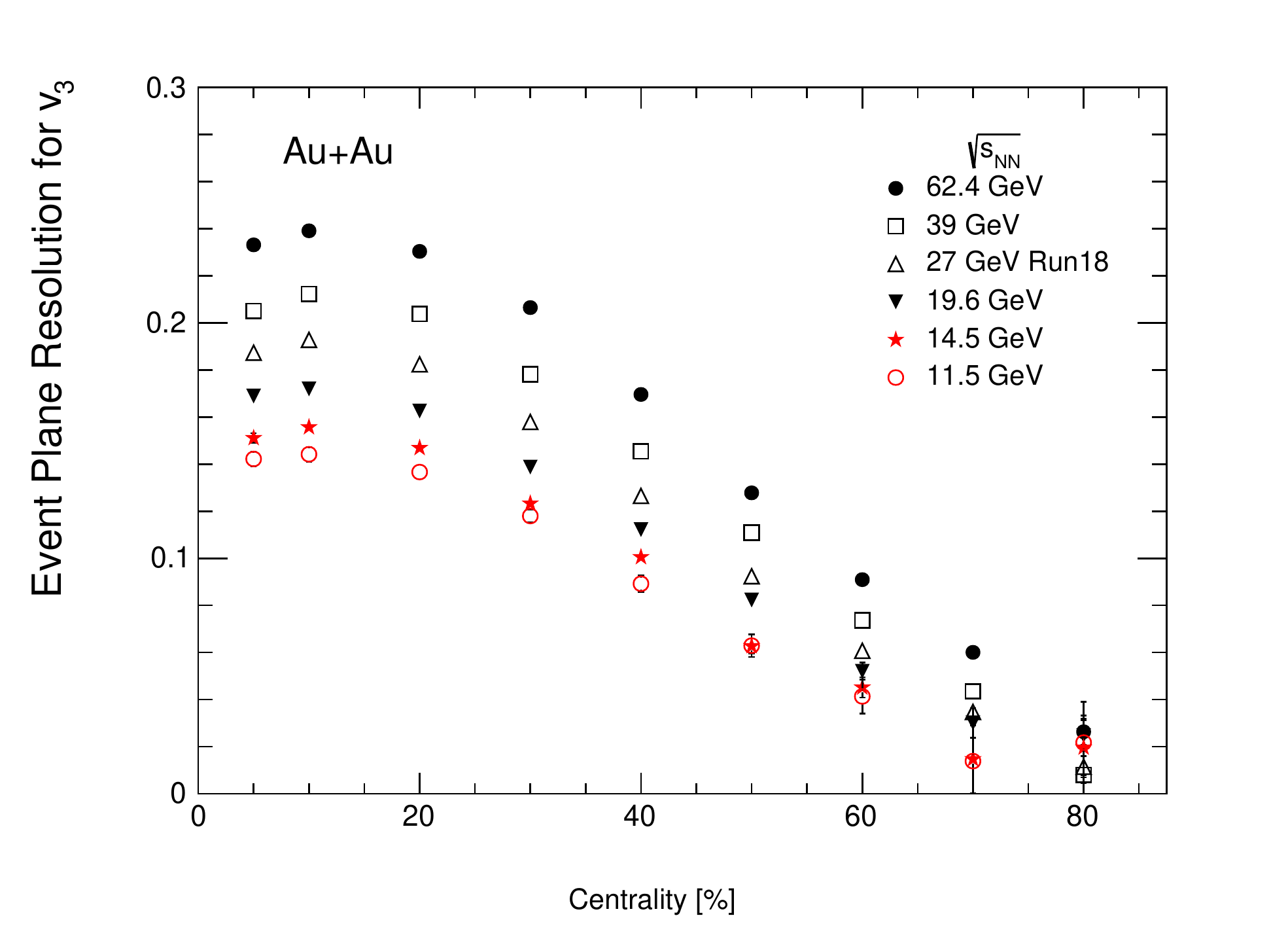}
	\caption{ The centrality dependence of the event plane resolution for $v_2$ (left panel) and $v_3$ (right panel)
         for all six collision energies.}
	\label{fig:res}
\end{figure}

Events were selected to have a vertex position about the nominal center of the TPC in the beam direction 
of $\pm$ 40 cm at $\mathrm{\sqrt{s_{_{NN}}}} = 62,~39,~27,~19.6$ and $14.5$~GeV, 
$\pm$ 50 cm at $\mathrm{\sqrt{s_{_{NN}}}} = 11.5$~GeV, and 
to be within a radius of $1-2$~cm with respect to the beam axis.  The centrality of each collisions was
determined by measuring event-by-event multiplicity and interpreting 
the measurement with a tuned Monte Carlo Glauber calculation~\cite{v2bes1,v2bes3}. Analyzed tracks were required to have
a distance of closest approach to the primary vertex to
be less than 3 cm, and to have at least 15 TPC space
points used in their reconstruction \cite{v2bes1,v2bes2}. The particle identification for
charged hadrons (~$\pi^\pm$, $K^\pm$, $p$, $\bar{p}$~) was based on a
combination of the ionization energy loss, $dE/dx$, in the TPC, and the squared mass,
$m^2$, from the TOF detector \cite{v2bes2,v2bes3,v2bes4}.\\
In this study, the event plane method with $\eta$ sub-events, separated
by an additional $\eta$-gap of $\Delta\eta >$ 0.1, was used to measure elliptic ($v_2$) and triangular ($v_3$)
flow ~\cite{v2bes1}. The $\eta$ gap is introduced to suppress the non-flow correlations between the two sub-events.
The $\eta$ sub-event method was implemented using the procedure in Ref~\cite{v2bes1,v2bes2,v2bes3}.
The centrality dependence of event plane resolution for $v_2$ and $v_3$ for
the six collision energies  is shown in the Fig.~\ref{fig:res}.  The systematic uncertainty
associated with the non-flow effects is estimated for each collision energy  by comparing $v_2$ and $v_3$
results obtained with different $\Delta\eta$ gaps. Studies were performed for $\Delta\eta$ values of 
0.1, 0.3, 0.5, 0.7.

\section{Results}

\begin{figure}[htb]
	\centering
	\includegraphics[width=0.43\textwidth]{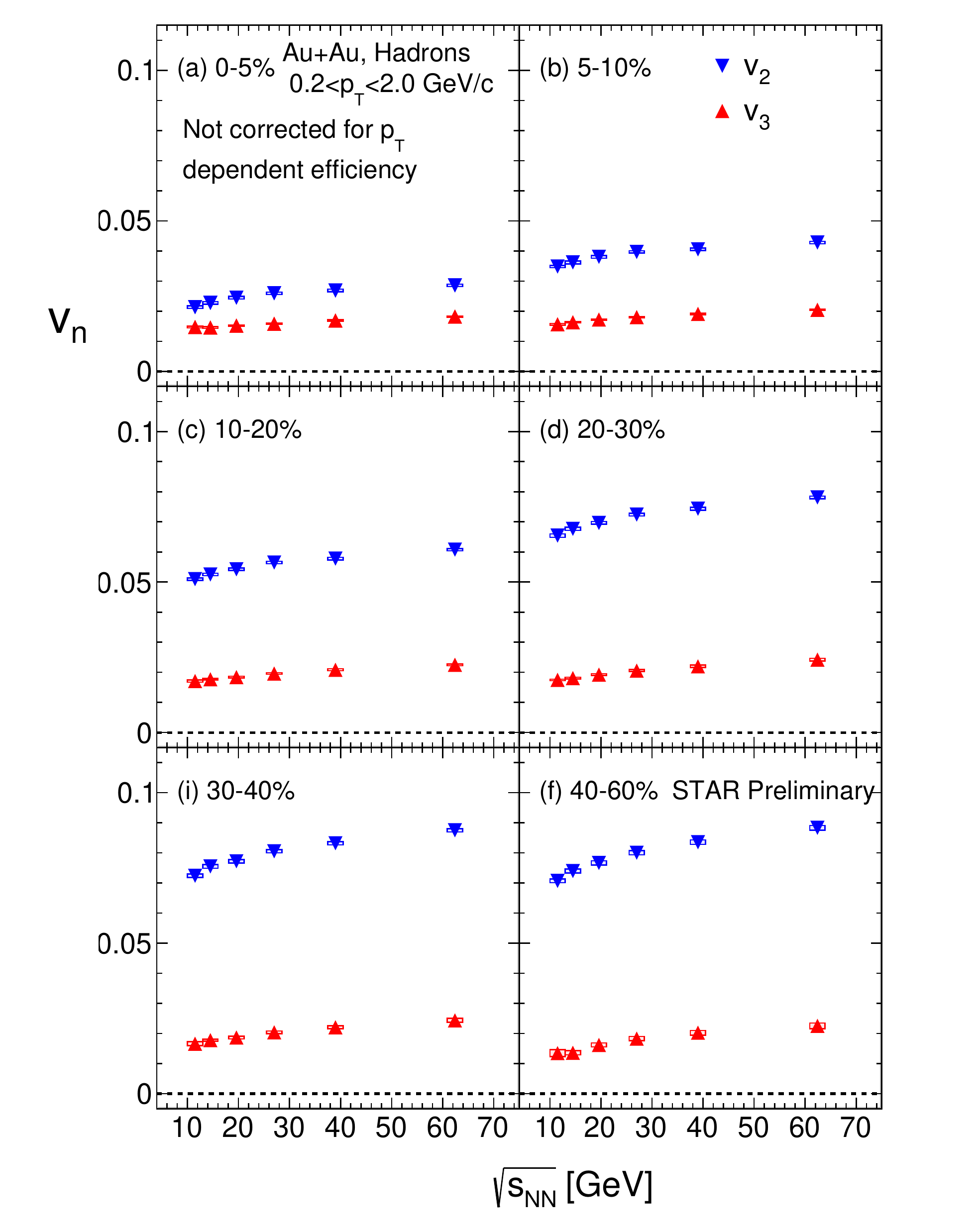}
	\includegraphics[width=0.43\textwidth]{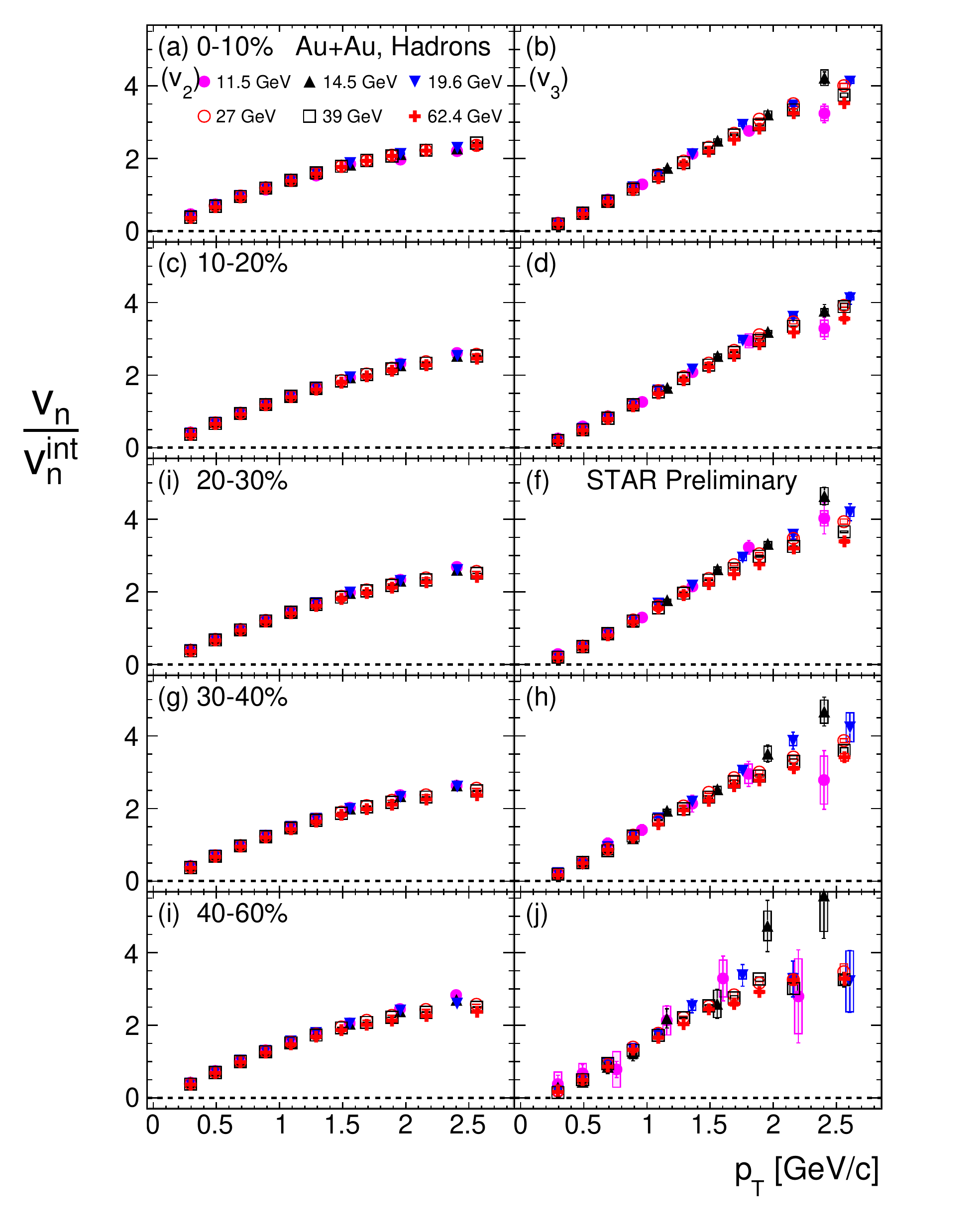}
	\caption{ Left: $p_T$-integrated $v_2^\textrm{int}$ and and $v_3^\textrm{int}$  of inclusive charged hadrons
          as a function of $\sqrt{s_{NN}}$ for different bins in collision centrality.
          Right: $p_T$-dependence of
          $v_2(p_T)/v_2^\textrm{int}$ and $v_3(p_T)/v_3^\textrm{int}$ of charged hadrons for different bins in collision centrality.
          The measured $v_n(p_T)$ values were divided by the corresponding  $v_n^\textrm{int}$ values from
          the left part of the figure. The results are presented for all 6 collision energies: $\sqrt{s_{NN}}$ = 11.5, 14.5, 19.6, 27, 39 and 62.4~GeV.}
	\label{fig:vn_chhad_vs_pT_BES}
\end{figure}
\begin{figure}[htb]
	\centering
	\includegraphics[width=0.43\textwidth]{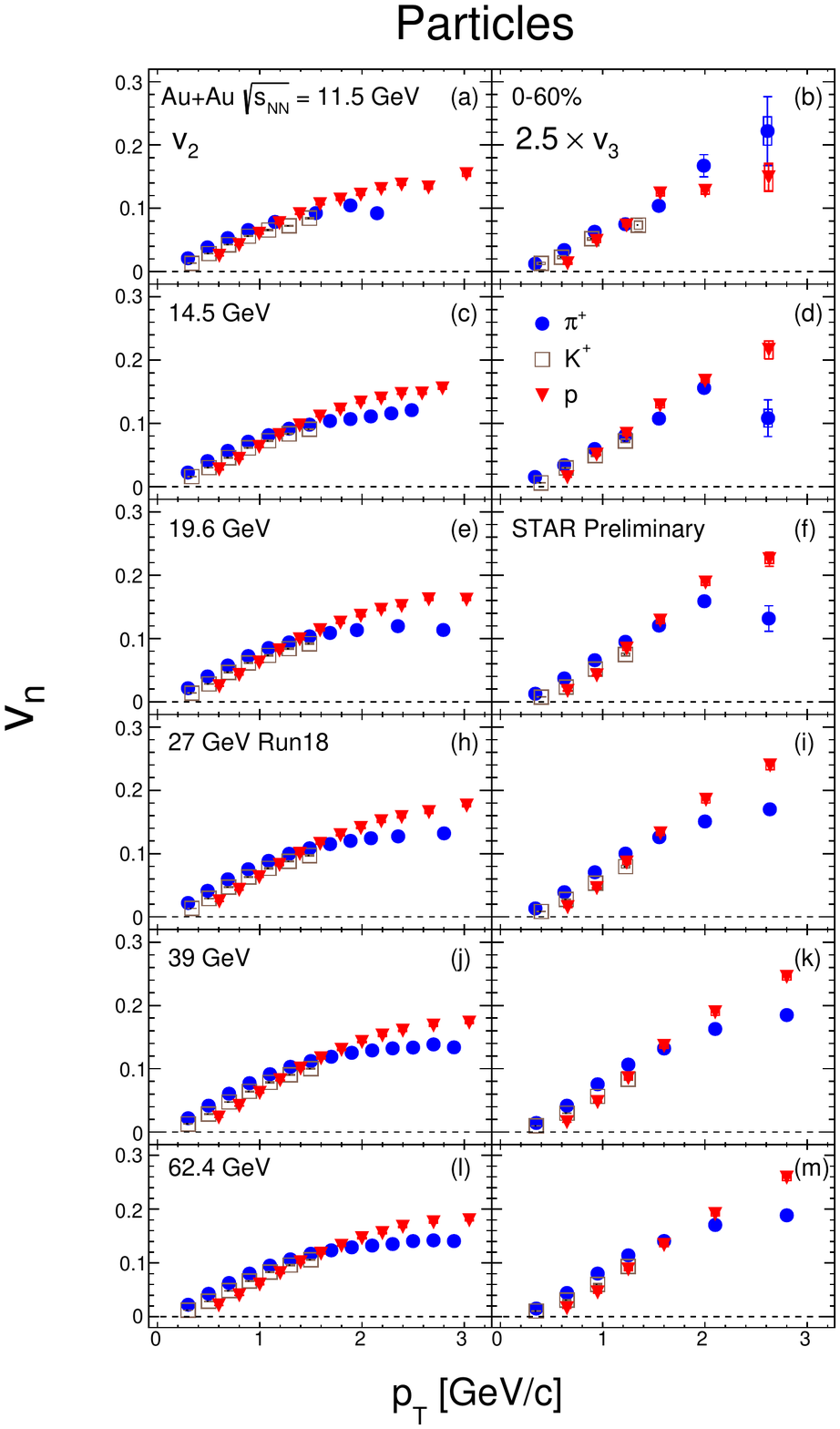}
	\includegraphics[width=0.43\textwidth]{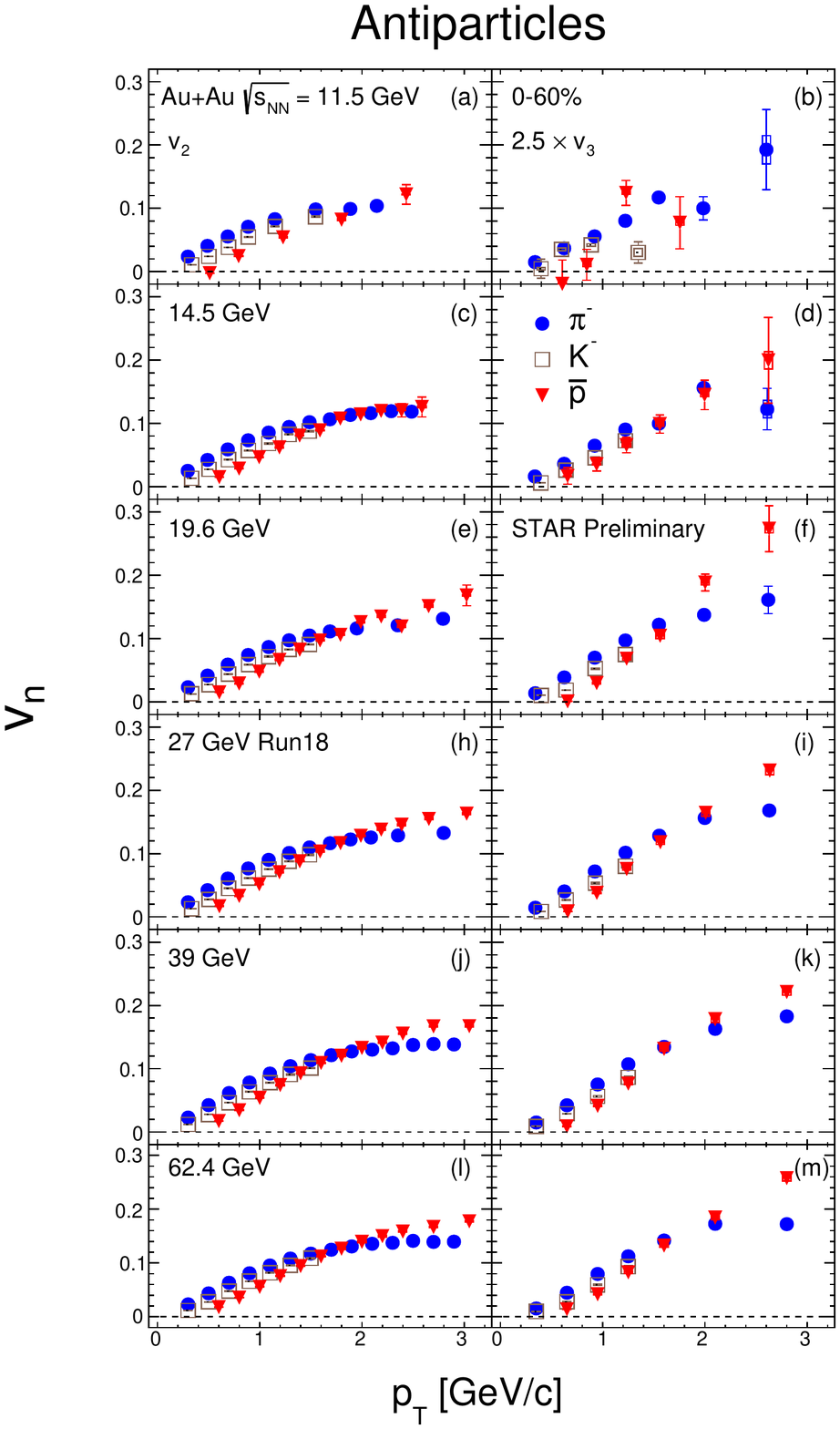}
	\caption{ $p_T$ dependence of $v_2$ and  $v_3$ signals of $\pi^+$, $K^+$, $p$ (left) and $\pi^-$, $K^-$, $\bar{p}$ (right)
          for 0-60\% central Au+Au collisions.}
	\label{fig:vn_vs_pT_1}
\end{figure}

Preliminary results for the excitation functions of  $p_T$-integrated (0.2 $< p_T <$ 3.2 GeV/c) 
values of $v_2^\textrm{int}$ and $v_3^\textrm{int}$ of  inclusive charged hadrons are presented in the left
part of Fig.~\ref{fig:vn_chhad_vs_pT_BES}. The 
$v_n^\textrm{int}$  results were not corrected for $p_T$ dependent tracking efficiency, which
will be explored in future analysis.
Although the efficiency is $p_T$ dependent but is similar between
different collision energy, so it is not expected to influence the $\sqrt{s_{NN}}$ trend.
The results are presented for 6 bins in collision centrality:
$0-5\%$, $5-10\%$, $10-20\%$, $20-30\%$, $30-40\%$ and $40-60\%$. The results indicate an
essentially monotonic increase for
$p_T$-integrated  $v_2$ and  $v_3$ with $\mathrm{\sqrt{s_{_{NN}}}}$, as expected from increase of the 
radial flow  with collision energy which pushes
the hadrons to larger $p_T$ and renders the momentum spectra less anisotropic at low $p_T$ \cite{v3bes}.
The $p_T$ dependence of  $v_2(p_T)/v_2^\textrm{int}$ and $v_3(p_T)/v_3^\textrm{int}$
 for inclusive charged hadrons is presented
in the right part of the Fig.~\ref{fig:vn_chhad_vs_pT_BES} for different bins in collision centrality.  The measured $v_n(p_T)$
values were divided by the corresponding  $p_T$-integrated $v_n^\textrm{int}$ values from
the left part of the Fig.~\ref{fig:vn_chhad_vs_pT_BES}. The results in the figure are presented for all 6 collision energies:
$\sqrt{s_{NN}}$ = 11.5, 14.5, 19.6, 27, 39 and 62.4~GeV and they show that $v_n(p_T)/v_n^\textrm{int}$
has a very week dependence on $\mathrm{\sqrt{s_{_{NN}}}}$. This is in agreement with
predictions from \cite{torr1}.\\
Figure~\ref{fig:vn_vs_pT_1} shows the collision energy dependence in $v_2(p_T)$ and $v_3(p_T)$ for 
identified hadrons (~$\pi^\pm$, $K^\pm$, $p$, $\bar{p}$~) for 0-60\% central Au+Au collisions. The
results for particles (left panel)  and anti-particles (right panel) are presented separately.
We observe that the  $v_3(p_T)$ signal  of identified charged hadrons exhibits similar trends as first observed for $v_2$ in
Au+Au collisions: mass  ordering at low transverse momenta, $p_T < 2$ GeV/c, and
meson/baryon splitting at intermediate $p_T$, $2< p_T < 4$ GeV/c \cite{v2bes2,v2bes3,v2bes4}.

\begin{figure}[htb]
	\centering
	\includegraphics[width=0.43\textwidth]{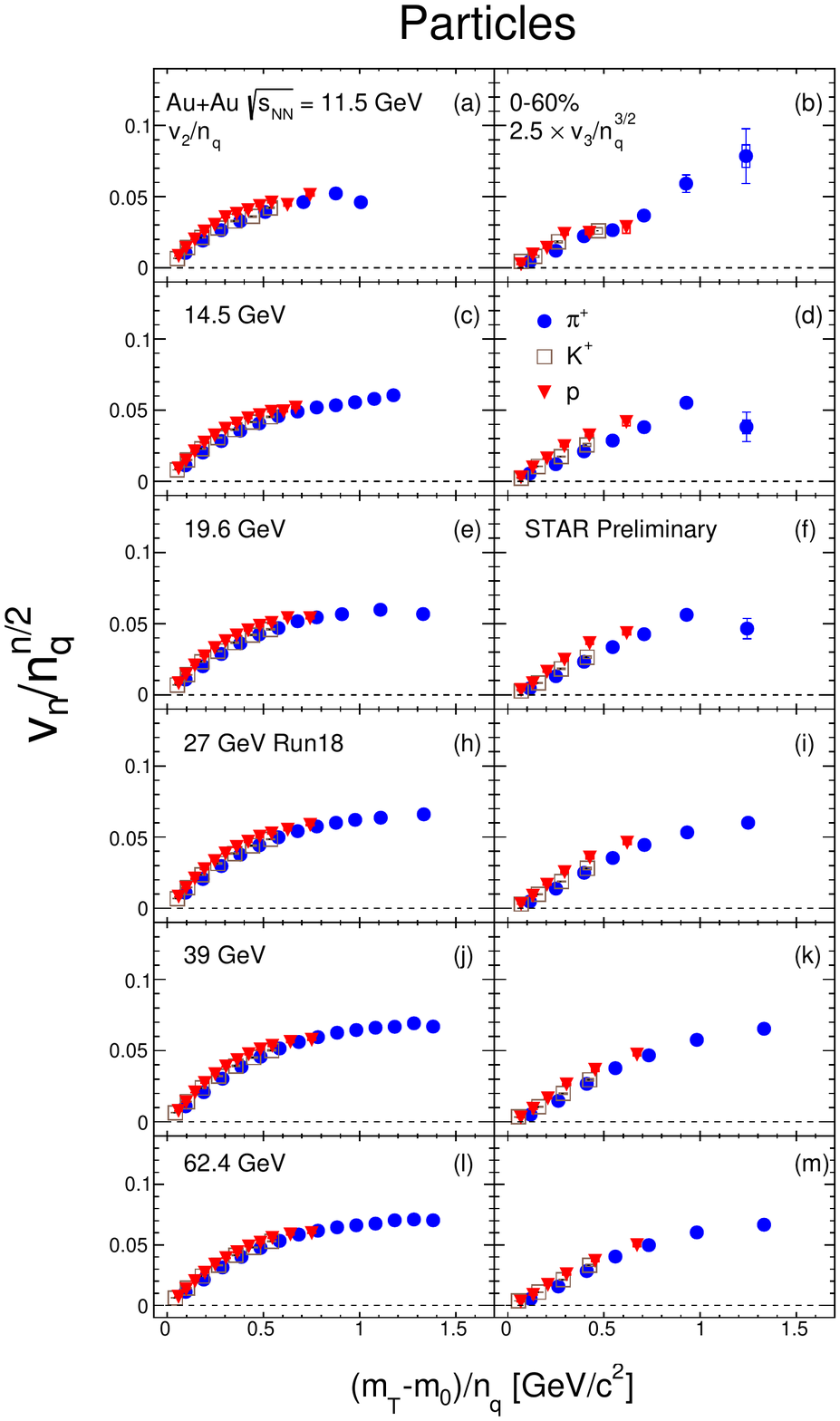}
	\includegraphics[width=0.43\textwidth]{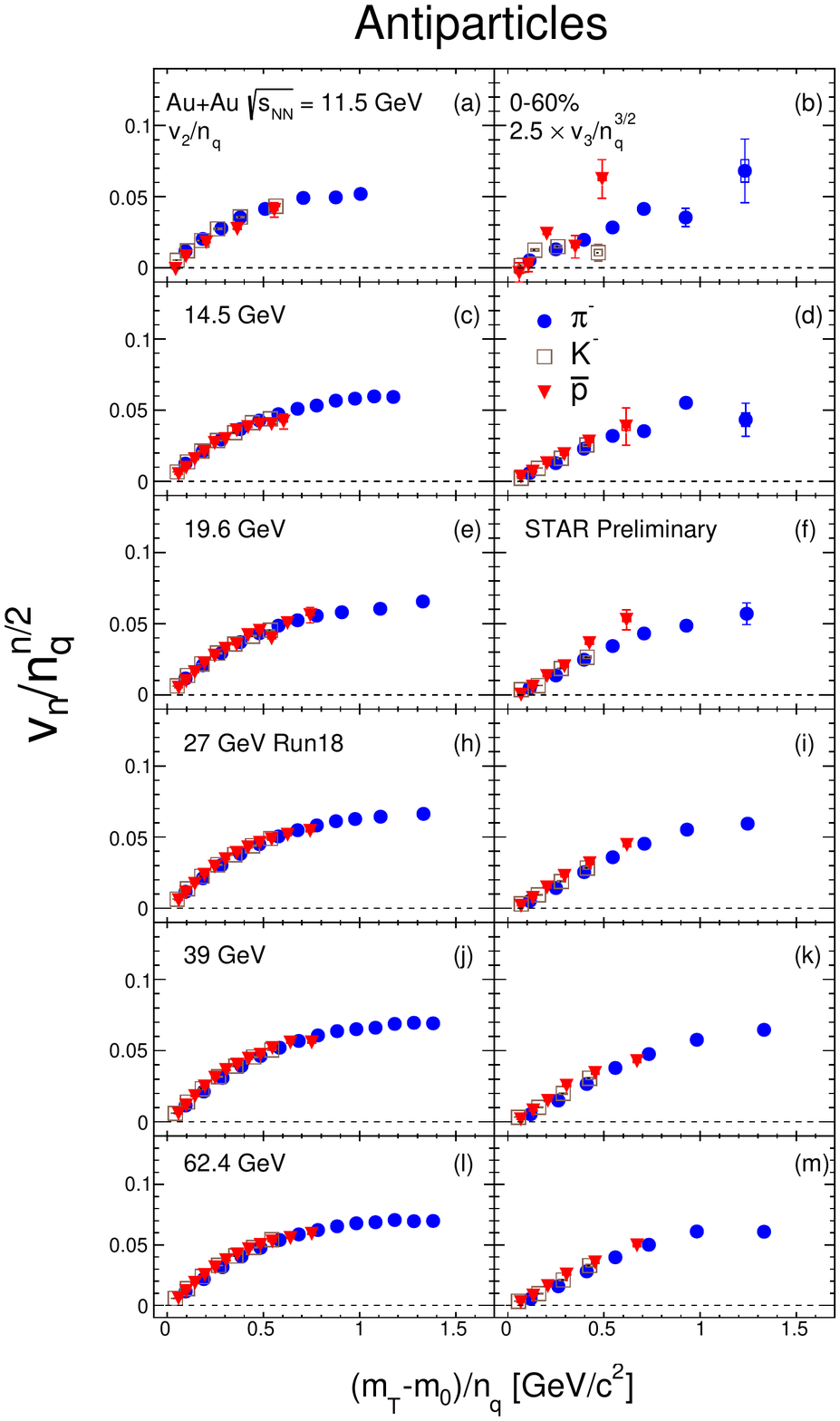}
	\caption{ The Number-of-Constituent Quark (NCQ) scaled elliptic and triangular flow,
          $v_n/n_q^{n/2}$ versus ($m_T-m_0$)/$n_q$, for 0-60\% central Au+Au collisions for
          for selected particles (left part)
          and corresponding anti-particles (right part).}
	\label{fig:vn_vs_pT_scale}
\end{figure}

\begin{figure}[htb]
	\centering
	\includegraphics[width=0.8\textwidth]{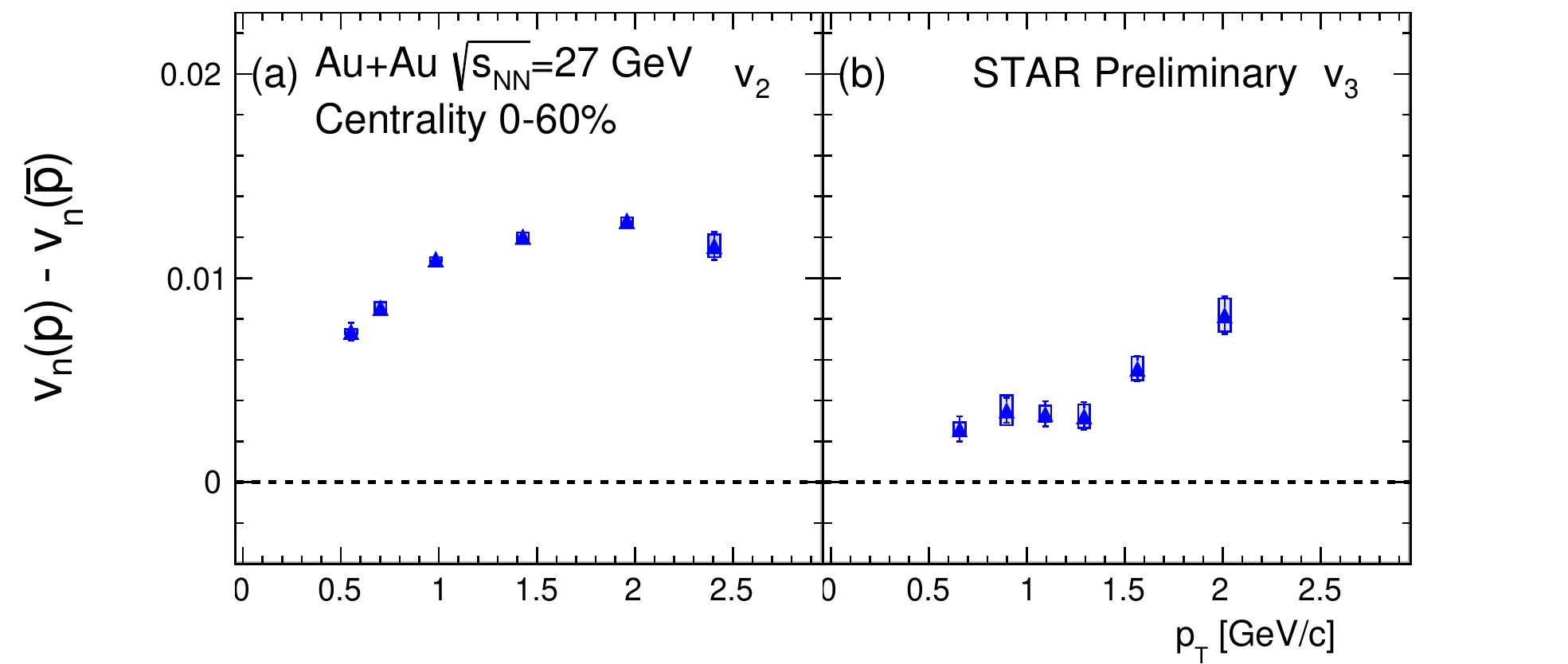}
	\caption{ The difference between proton and antiproton $v_n$ as a function of the transverse momentum
          $p_T$ for 0-60\% central Au+Au collisions at $\sqrt{s_{NN}}=27$~GeV.}
	\label{fig:vn_Diff_vs_pT_27gev}
\end{figure}

Figure.~\ref{fig:vn_vs_pT_scale} shows that the measured $v_3$ values of identified charged hadrons seems to follow
the NCQ scaling, $v_n/n_q^{n/2}$ versus ($m_T-m_0$)/$n_q$, if we plot the results
for particles (left part) and anti-particles (right part) separately.\\
The analysis of
the new dataset of Au+Au collisions at $\sqrt{s_{NN}}=27$~GeV, collected by STAR experiment in 2018, allows us to
observe the difference in the triangular $v_3$ flow  between protons and anti-protons, see Fig.~\ref{fig:vn_Diff_vs_pT_27gev}.
It shows that, similar to elliptic flow $v_2$, the $v_3$  flow signal  of protons is larger
than $v_3$ of antiprotons and the difference has $p_T$ dependence.

\begin{figure}[htb]
	\centering
	\includegraphics[width=0.37\textwidth]{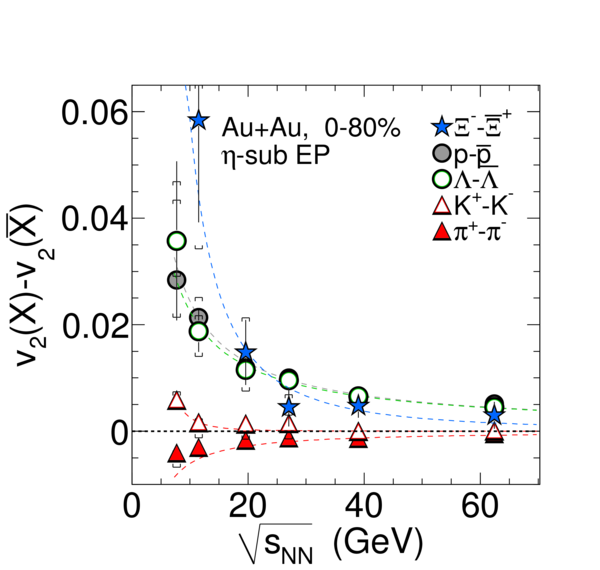}
	\includegraphics[width=0.40\textwidth]{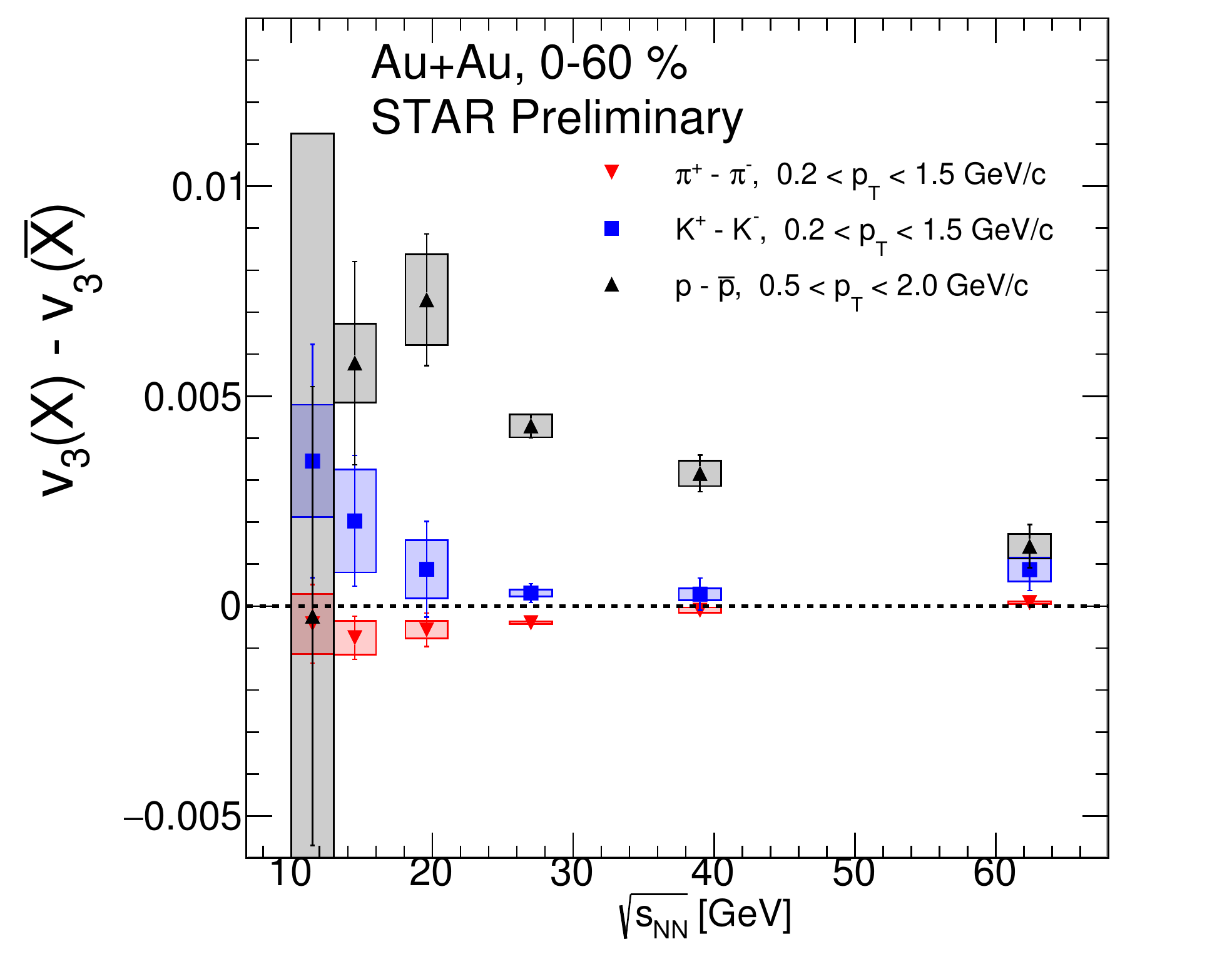}
	\caption{ (left) The difference in $v_2$ between particles (X) and their corresponding anti-particles ($\bar{X}$)
          (see legend) as a function of $\sqrt{s_{NN}}$ for $0-80\%$ central Au+Au collisions. The figure
          is taken from \cite{v2bes3}.
          (right) The preliminary results for the
          difference in the $v_3$ values between particles ($X$) and its corresponding anti-particles
          ($\bar{X}$) as a function of  $\sqrt{s_{NN}}$ for 0-60\% central Au+Au collisions.}
	\label{fig:v3_Diff_vs_energy}
\end{figure}

For other collision energies we can estimate the difference in $v_3$ values between particles and corresponding
anti-particles for $p_T$-integrated $v_3$ values.
The right part of Fig.~\ref{fig:v3_Diff_vs_energy} shows the difference in $v_3$
between particles (X) and their corresponding anti-particles ($\bar{X}$)
as a function of $\sqrt{s_{NN}}$ for $0-60\%$ central Au+Au collisions. Similar to $v_2$, the $v_3(X)-v_3(\bar{X})$
difference increases with decreasing collision energy and it is larger for baryons than mesons \cite{v2bes2,v2bes3,v2bes4}.

\section{Summary}
In summary, we have employed the event plane method with $\eta$ sub-events
to carry out new measurements of the  triangular ($v_3$) anisotropic flow coefficients for inclusive and identified
charged hadrons (~$\pi^\pm$, $K^\pm$, $p$, $\bar{p}$~) at midrapidity in Au+Au collisions, spanning the
collision energy range $\sqrt{s_{NN}}$ = $11.5$ - $62.4$~GeV.  We observe that the triangular flow
signal ($v_3$) of identified hadrons exhibits similar trends as first observed
for $v_2$ \cite{v2bes1,v2bes2,v2bes3}. New measurements of $v_3$ excitation function  
could serve as constraints to test  different models and to aid new information about the
temperature dependence of the transport properties of the strongly interacting matter.

\section{Acknowledgments}
This work is supported by the RFBR according to the research project No. 18-02-40086 and
by the Ministry of Science and Higher Education of the Russian Federation,
Project "Fundamental properties of elementary particles and cosmology" No 0723-2020-0041.


\section*{References}
\begin{thebibliography}{9}
\bibitem{starQM}
  J.~Adams {\it et al.} [STAR Collaboration],
  Nucl.\ Phys.\ A {\bf 757} (2005) 102

\bibitem{phenixQM}
  K.~Adcox {\it et al.} [PHENIX Collaboration],
  Nucl.\ Phys.\ A {\bf 757} (2005) 184

\bibitem{bass}
J.~E.~Bernhard, J.~S.~Moreland and S.~A.~Bass,
Nature Phys. \textbf{15} (2019) no.11, 1113-1117


\bibitem{lqcd}
  Y.~Aoki, G.~Endrodi, Z.~Fodor, S.~D.~Katz and K.~K.~Szabo,
  Nature {\bf 443} (2006) 675

\bibitem{starbes1}
  D.~Keane,
  J.\ Phys.\ Conf.\ Ser.\  {\bf 878} (2017) no.1,  012015.


\bibitem{starbes2} H.~Caines,
  Nucl.\ Phys.\ A {\bf 967} (2017) 121.

\bibitem{vol3}
S.~A.~Voloshin, A.~M.~Poskanzer and R.~Snellings,
Landolt-Bornstein \textbf{23} (2010), 293-333

\bibitem{lhc2}
  R.~Snellings,
  J.\ Phys.\ G {\bf 41} (2014) no.12,  124007

\bibitem{vol1}
  S.~Voloshin and Y.~Zhang,
  Z.\ Phys.\ C {\bf 70} (1996) 665



\bibitem{v2v3phenix}
A.~Adare \textit{et al.} [PHENIX],
Phys. Rev. C \textbf{93} (2016) no.5, 051902

\bibitem{v2v3star}
L.~Adamczyk \textit{et al.} [STAR],
Phys. Rev. C \textbf{98} (2018) no.1, 014915


\bibitem{v2v3alice}
J.~Adam \textit{et al.} [ALICE],
JHEP \textbf{09} (2016), 164

\bibitem{v3hydro} B.~Schenke, S.~Jeon and C.~Gale,
Phys. Rev. Lett. \textbf{106} (2011), 042301

\bibitem{v3hybrid}
  J.~Auvinen and H.~Petersen,
  Phys.\ Rev.\ C {\bf 88} (2013) no.6,  064908

 \bibitem{v2D}
L.~Adamczyk \textit{et al.} [STAR],
Phys. Rev. Lett. \textbf{118} (2017) no.21, 212301

\bibitem{v2bes1}
L.~Adamczyk \textit{et al.} [STAR],
Phys. Rev. C \textbf{86}, 054908 (2012)

\bibitem{v2bes2}
L.~Adamczyk \textit{et al.} [STAR],
Phys. Rev. C \textbf{88}, 014902 (2013)

\bibitem{v2bes3}
L.~Adamczyk \textit{et al.} [STAR],
Phys. Rev. Lett. \textbf{110} (2013) no.14, 142301


\bibitem{v2bes4}
L.~Adamczyk \textit{et al.} [STAR],
Phys. Rev. C \textbf{93}, no.1, 014907 (2016)


\bibitem{v3bes}
L.~Adamczyk \textit{et al.} [STAR],
Phys. Rev. Lett. \textbf{116} (2016) no.11, 112302

\bibitem{torr1}
G.~Torrieri,
Eur. Phys. J. A \textbf{52} (2016) no.8, 249



\end{thebibliography}
\end{document}